\documentclass[12pt]{article}
\usepackage{amsmath}
\usepackage{amsfonts}
\usepackage{amssymb}
\usepackage{url}
\usepackage{color}
\usepackage{graphicx}
\newcommand{\be}{\begin{equation}}
\newcommand{\ee}{\end{equation}}
\newcommand{\bea}{\begin{eqnarray}}
\newcommand{\eea}{\end{eqnarray}}
\newcommand{\bref}[1]{(\ref{#1})}
\begin{document}
\begin{titlepage}
\begin{flushright}
\today
\end{flushright}
\vspace{4\baselineskip}
\begin{center}
{\Large\bf  Zero Texture Model and SO(10) GUT}
\end{center}
\vspace{1cm}
\begin{center}
{\large Takeshi Fukuyama$^{a,}$
\footnote{E-mail:fukuyama@se.ritsumei.ac.jp}},
{\large Koichi Matsuda$^{b,}$
\footnote{E-mail:matsuda@mail.tsinghua.edu.cn}}
and
{\large Hiroyuki Nishiura$^{c,}$
\footnote{E-mail:nishiura@is.oit.ac.jp}}
\end{center}
\vspace{0.2cm}
\begin{center}
${}^{a}$ {\small \it Department of Physics, Ritsumeikan University,
Kusatsu, Shiga, 525-8577, Japan}\\[.2cm]
${}^{b} $ {\small \it Center for High Energy Physics, Tsinghua University, Beijing 100084, China}\\[.2cm]
${}^{c} $ {\small \it Faculty of Information Science and Technology, Osaka Institute of Technology, Hirakata, Osaka 573-0196, Japan}
\medskip
\vskip 10mm
\end{center}
\vskip 10mm
\begin{abstract}
We study the bridge between the phenomenological mass matrix model and SO(10) GUT. 
Namely, we consider the four zero texture model in the framework of the renormalizable SO(10) GUT model. 
This unification gives more stringent constraints than the case where only either model is considered. 
However, we can obtain good fitting by expanding the minimal SO(10) GUT to include ${\bf 120}$ in addition to ${\bf 10}$ and $\overline{{\bf 126}}$ in Yukawa coupling and by considering both type-I and type-II seesaw mechanisms. %
\end{abstract}
\end{titlepage}
\section{Introduction}
Many papers have been published on the phenomenological mass matrix model of quark-lepton \cite{shizuoka}. 
In this approach flavour symmetry and texture zero are important ingredients. 
In most of these models, quark and lepton are discussed on the same footing. 
This approach is rather directly related with observations. 
However there has no base why and how quark and lepton are unified.
The latter is the main theme of the grand unified model, 
which can say the relations between mass matrices but not on the mass matrix itself.
So these two approaches should be complementary to each other.
However, curiously enough, such attempts have been rather few.
\cite{nishiura3} are those of such rare attempts, in which we discussed 
the relation between four  zero texture (FZT) \footnote{In the recent 
papers there is a confusion on this terminology. 
We have called this model FZT since we consider up and down quark mass matrices (or lepton doublet) as a unit and they have totally 2+2=4 zeros.} mass matrix model and 
renormalizable SO(10) GUT \cite{fukuyama}.
Renormalizable SO(10) GUT implies that fermion-Higgs coupling is resticted in 
the renormalizable Yukawa coupling.
The fundamental representation of matter multiplets are ${\bf 16}_i$ 
(i=family index), and
${\bf 16}\times {\bf 16}={\bf 10}+{\bf 120}+{\bf 126}$.
So the general Yukawa coupling is given by 
\be
Y_{10,ij}{\bf 16}_i{\bf 16}_j{\bf 10}_H+Y_{120,ij}{\bf 16}_i{\bf 16}_j{\bf 120}_H+Y_{126,ij}{\bf 16}_i{\bf 16}_j\overline{{\bf 126}}_H \ .
\ee
In this paper we combine this model with FZT model.
On the side of SO(10) we consider the most general model with use of both type I and type II seesaw mechanism 
for neutrino masses.
\section{Four Zero texture Quark-Lepton Mass Matrices in SO(10) GUT}
Phenomenological quark mass matrices have been discussed from various 
points of view \cite{shizuoka}.  In this section we review our quark and lepton mass matrix model \cite{nishiura3}.  
The mass term in the Lagrangian is given by 
\begin{eqnarray}
L_M = 
-\overline{q_{R,i}^{u}}M_{uij}q_{L,j}^{u}
-\overline{q_{R,i}^{d}}M_{dij}q_{L,j}^{d}
-\overline{l_{R,i}}M_{eij}l_{L,j} 
-\overline{\nu'_{R,i}}M_{Dij}\nu_{L,j} \nonumber \\
-\frac{1}{2}\overline{(\nu_{L,i})^{c}}M_{Lij}\nu_{L,j}
-\frac{1}{2}\overline{(\nu'_{R,i})^{c}}M_{Rij}\nu'_{R,j}+H.c.
\end{eqnarray}
with
\begin{equation}
q_{L,R}^{u}=
\left(
	\begin{array}{c}
	u \\
	c \\
	t
	\end{array}
\right)_{L,R}, \quad
q_{L,R}^{d}=
\left(
	\begin{array}{c}
	d \\
	s \\
	b
	\end{array}
\right)_{L,R}, \quad
l_{L,R}=
\left(
	\begin{array}{c}
	e \\
	\mu \\
	\tau
	\end{array}
\right)_{L,R}, \quad
\nu_{L}=
\left(
	\begin{array}{c}
	\nu_e \\
	\nu_{\mu} \\
	\nu_{\tau}
	\end{array}
\right)_{L}, \quad
\nu'_{R}=
\left(
	\begin{array}{c}
	\nu'_e \\
	\nu'_{\mu} \\
	\nu'_{\tau}
	\end{array}
\right)_{R},
\end{equation}
where 
$M_u$, $M_d$, $M_D$, $M_e$,$M_L$, and $M_R$ are the mass matrices for up quarks, 
down quarks, Dirac neutrinos, charged leptons, left-handed Majorana neutrinos, and right-handed Majorana neutrinos, respectively.
The mass matrix of light Majorana neutrinos $M_\nu$ is given by 
\begin{equation}
M_\nu=M_L-M_D^T M_R^{-1} M_D, \label{eq051102}
\end{equation}
which is constructed via the seesaw mechanism 
\cite{yanagida} 
from the block-diagonalization of neutrino mass matrix,
\begin{equation}
\left(
        \begin{array}{cc}
        M_L & M_D^T \\
        M_D & M_R 
        \end{array}
\right).
\end{equation}
\par
In the SO(10) GUT scheme, we consider the general renormalizable model, in whichYukawa coupling involves three types of mass matrices involving ${\bf 10},~{\bf 120},$ and ${\bf \overline{126}}$ Higgs. 
These Higgs fields are decomposed into
\bea
{\bf 10}&=&(\bf{6,1,1})+({\bf 1,2,2}),\nonumber\\
{\bf 120}&=&({\bf 15,2,2})+({\bf 6,3,1})+({\bf 6,1,3})+({\bf 1,2,2})+({\bf 10,1,1})+({\bf \overline{10},1,1}),\\
{\bf 126}&=&({\bf 10,1,3})+({\bf \overline{10},3,1})+({\bf 15,2,2})+({\bf 6,1,1}).\nonumber
\eea
under $SU(4)\times SU(2)_L\times SU(2)_R$.
Thus ${\bf 10}$ and ${\bf 126}$ have two SU(2) doublets and appears CG coefficients 1, -3, respectively. On the other hand, ${\bf 120}$ has four doublets and no CG coefficient. 
And
\bea
{\bf 16}\times {\bf 16}\times {\bf 10}&\supset& {\bf 5}_H(uu^c+\nu\nu^c)+{\bf \overline{5}}_H(dd^c+ee^c),\nonumber\\
{\bf 16}\times {\bf 16}\times {\bf 120}&\supset& {\bf 5}_H\nu\nu^c+{\bf 45}_Huu^c+{\bf \overline{5}}_H(dd^c+ee^c)+{\bf \overline{45}}_H(dd^c-3ee^c),\\
{\bf 16}\times {\bf 16}\times {\bf \overline{126}}&\supset& {\bf 1}_H\nu^c\nu^c+{\bf 15}_H\nu\nu+{\bf 5}_H(uu^c-3\nu\nu^c)+{\bf \overline{45}}_H(dd^c-3ee^c).\nonumber
\eea
under SU(5) decomposition. The first and second terms of $\overline{{\bf 126}}$ are the right-handed and left-handed Majorana masses, respectively. So six mass matrices included in SO(10) have the following form,
\bea
M_u&=&S+\delta'A+\epsilon S'\equiv S_u + A_u,\nonumber\\
M_d&=&\alpha S+\delta A+S'\equiv S_d + A_d,\nonumber\\
M_D&=&S+\delta'' A-3\epsilon S'\equiv S_D + A_D,\\
M_e&=&\alpha S+A-3S'\equiv S_e + A_e,\nonumber\\
M_L&=&~~~~~~~~~~~~~\beta S'\equiv S_L,\nonumber\\
M_R&=&~~~~~~~~~~~~~\gamma S'\equiv S_R.\nonumber
\label{mass form}
\eea
Here $S,A,S'$ represent common structure for mass matrices 
which come from ${\bf 10}$, ${\bf 120}$, and ${\bf 126}$,  
respectively.
$\alpha,\delta, \delta', \delta'',\epsilon, \beta, \gamma$  are relative coefficients of vacuum expectation values (VEVs).
$S_f(A_f)$ ($f=u,d,D,e,L,R$) are the symmetric (antisymmetric) part of $M_f$.
Furthermore,we put a ansatz that the mass matrices $M_{u}, M_{d}, M_{e}$ and $M_{\nu}$ 
are hermitian and have 
the same textures.  That is, $S_\alpha$ are real and $A_\alpha$ are pure imaginary. Our model is  different from the Fritzsch model \cite{fritzsch} in the 
sense that (2,2) components are not zeros and that our model deals with 
the quark and lepton mass matrices on the same footing.  
The mass matrices \(M_D\), \(M_L\), and \(M_R\) are, furthermore, assumed 
to have the same zero 
texture as \(M_\nu\) \cite{matsuda}.  This ansatz restricts the texture 
forms \cite{nishiura3} and we 
choose 
the following our texture because it is most closely related with the NNI 
form \cite{branco}.
\begin{equation}
\mbox{NNI}:
\left(
        \begin{array}{ccc}
         0  & * &  0 \\
         {*}  & 0 &  *\\
         0  & * &  *
        \end{array}
\right), \hspace{3cm}
\mbox{{\bf Our Texture}}:
\left(
        \begin{array}{ccc}
         0  & * &  0 \\
         {*}  & * &  *\\
         0  & * &  *
        \end{array}
\right).
\end{equation}
Our model is  different from the NNI model in the 
sense that (2,2) components are not zeros and that our model deals with 
the quark and lepton mass matrices on the same footing.  
Thus, in the FZT model, the quark and lepton mass matrices are described as follows.
\begin{eqnarray}
&&M_{u}=
P_u\left(
        \begin{array}{ccc}
         0    & a_{u} & 0 \\
        a_{u} & b_{u} & c_{u}\\
         0    & c_{u} & d_{u}
        \end{array}
\right)P_u^\dagger
=
\left(
        \begin{array}{ccc}
         0    & a_{u}e^{i\tau_{u}} & 0 \\
        a_{u}e^{-i\tau_{u}} & b_{u} & c_{u}e^{i\sigma_{u}}\\
         0    & c_{u}e^{-i\sigma_{u}} & d_{u}
        \end{array}
\right), \nonumber \\
&&M_{d}=
P_d \left(
        \begin{array}{ccc}
         0  & a_{d} &  0 \\
        a_{d} & b_{d} & c_{d}\\
         0  & c_{d} & d_{d}
        \end{array}
\right) P_d^\dagger 
=
\left(
        \begin{array}{ccc}
         0  & a_{d} e^{i\tau_{d}} &  0 \\
        a_{d} e^{-i\tau_{d}} & b_{d} & c_{d} e^{i\sigma_{d}}\\
         0  & c_{d} e^{-i\sigma_{d}} & d_{d}
        \end{array}
\right),\nonumber\\
&&M_{e}=
P_e \left(
        \begin{array}{ccc}
         0  & a_{e} &  0 \\
        a_{e} & b_{e} & c_{e}\\
         0  & c_{e} & d_{e}
        \end{array}
\right) P_e^\dagger 
=
\left(
        \begin{array}{ccc}
         0  & a_{e} e^{i\tau_{e}} &  0 \\
        a_{e} e^{-i\tau_{e}} & b_{e} & c_{e} e^{i\sigma_{e}}\\
         0  & c_{e} e^{-i\sigma_{e}} & d_{e}
        \end{array}
\right),\label{eq051101}\\
&&M_{L}=
\left(
        \begin{array}{ccc}
         0  & a_{L} &  0 \\
        a_{L} & b_{L} & c_{L}\\
         0  & c_{L} & d_{L}
        \end{array}
\right), \nonumber \\
&&M_{R}=
\left(
        \begin{array}{ccc}
         0  & a_{R} &  0 \\
        a_{R} & b_{R} & c_{R}\\
         0  & c_{R} & d_{R}
        \end{array}
\right), \nonumber 
\end{eqnarray}
where 
\(P_f\equiv\mbox{diag}(e^{i\gamma_{f1}},e^{i\gamma_{f2}},e^{i\gamma_{f3}})\),
\(\tau_{f}\equiv \gamma_{f1}-\gamma_{f2}\) and \(\sigma_{f}\equiv \gamma_{f2}-\gamma_{f3}\).
Hereafter we collectively describe these matrices as
\be
M_f=
\left(
        \begin{array}{ccc}
         0  & a_{f} e^{i\tau_f} &  0 \\
        a_{e} e^{-i\tau_f} & b_{f} & c_{f} e^{i\sigma_f}\\
         0  & c_{f} e^{-i\sigma_f} & d_{f}
        \end{array}
\right)
\ee
in addition to \bref{mass form}. 
Here f=u,d,D,e,L,R. 
If we denote the mass eigen values of $M_f$ as $m_{fi}~(i=1,2,3)$,
then $a_f,~b_f,~c_f$ can be expressed in terms of $m_{fi}$ and $d_f$:
\begin{align}
a_f&=\sqrt{-\frac{m_{f1}m_{f2}m_{f3}}{d_f}}, &
c_f&=\sqrt{-\frac{(d_f-m_{f1})(d_f-m_{f2})(d_f-m_{f3})}{d_f}},\nonumber\\
b_f&=m_{f1}+m_{f2}+m_{f3}-d_f.\label{abc}
\end{align}
Here 
\begin{align}
0&<m_{f1}<-m_{f2}<m_{f3} & \mbox{for }&|m_{f1}|<d_f<|m_{f2}|,\nonumber\\
0&<-m_{f1}<m_{f2}<m_{f3} & \mbox{for }&|m_{f2}|<d_f<|m_{f3}|.
\end{align}
The number of parameters of quark and charged lepton are from f=u,d,e of \bref{mass form},
\be
S,A,S'\rightarrow 4+2+4=10
\ee
and 4 coefficients of VEVs, $\alpha,\delta, \delta', \epsilon$, and totally $10+4=14$. On the other hand, the number of constraints from experiments are
quark masses (6), CKM mixing angles (3), the Dirac phase (1) and the lepton masses (3), that is, $6+3+1+3=13$.
Therefore only 1 parameter is remained after the data fitting of quark and charged lepton.
Using this one free parameter and $\delta'',\beta,\gamma$, we can fix light neutrino masses (3) and MNS mixing angles (3) and the phases (3).
Among these the direct experimental constraints are $\Delta^2 m_{23},\Delta^2 m_{12}$
and three MNS mixing angles.
\section{Numerical analyses}
FZT matrices are diagonalized as
\be
U_f^\dagger M_fU_f=\mbox{diag}(m_{f1},m_{f2},m_{f3}).
\ee
Here
\be
U_f=P_f^\dagger O_f,~~P_f=\mbox{diag}(1,\tau_f,\sigma_f+\tau_f)\equiv (1,\alpha_{f2},\alpha_{f3}).
\ee
and
\be
O_f=
\left(
        \begin{array}{ccc}
         \sqrt{\frac{(d_f-m_{f1})m_{f2}m_{f3}}{R_{f1}d_f}},  & \sqrt{\frac{(d_f-m_{f2})m_{f1}m_{f3}}{R_{f2}d_f}}, &  \sqrt{\frac{(d_f-m_{f3})m_{f2}m_{f1}}{R_{f3}d_f}} \\
        -\sqrt{-\frac{(d_f-m_{f1})m_{f1}}{R_{f1}}}, & \sqrt{-\frac{(d_f-m_{f2})m_{f2}}{R_{f2}}}, & \sqrt{-\frac{(d_f-m_{f3})m_{f3}}{R_{f3}}}\\
         \sqrt{\frac{m_{f1}(d_f-m_{f2})(d_f-m_{f3})}{R_{f1}d_f}},  & -\sqrt{\frac{m_{f2}(d_f-m_{f3})(d_f-m_{f1})}{R_{f2}d_f}}, & \sqrt{\frac{m_{f3}(d_f-m_{f1})(d_f-m_{f2})}{R_{f3}d_f}}
        \end{array}
\right),
\ee
where
\bea
R_{f1}&\equiv& (m_{f1}-m_{f2})(m_{f1}-m_{f3}),~R_{f2}\equiv (m_{f2}-m_{f3})(m_{f2}-m_{f1}),\nonumber\\
R_{f3}&\equiv&(m_{f3}-m_{f1})(m_{f3}-m_{f2}).
\eea
Hereafter we denote $(m_{f1}, m_{f2}, m_{f3})$ for $f=u, d,$ and $e$ as
$(m_{u}, m_{c}, m_{t})$, $(m_{d}, m_{s}, m_{b})$, and $(m_{e}, m_{\mu}, m_{\tau})$, 
respectively.
The CKM quark mixing matrix $U_{CKM}\equiv U_u^\dagger U_d$ is given by
\bea
(U_{CKM})_{12}&\approx& \sqrt{\frac{|m_d|}{m_s}}-e^{i\alpha_2}\sqrt{\frac{|m_u|}{m_c}x_ux_d}-e^{i\alpha_3}\sqrt{\frac{|m_u|}{m_c}(1-x_u)(1-x_d)}\, ,\nonumber\\
(U_{CKM})_{23}&\approx& \sqrt{\frac{|m_d|m_s}{m_b^2}}-e^{i\alpha_2}\sqrt{\frac{|m_u|}{m_c}x_u(1-x_d)}+e^{i\alpha_3}\sqrt{\frac{|m_u|}{m_c}(1-x_u)x_d}\, ,\label{CKMexp}\\
(U_{CKM})_{13}&\approx& \sqrt{\frac{|m_u||m_d|m_s(1-x_d)}{m_cm_b^2x_d}}+e^{i\alpha_2}\sqrt{x_u(1-x_d)}-e^{i\alpha_3}\sqrt{(1-x_u)x_d}\, .\nonumber
\eea
and the Dirac phase is
\be
\delta_q
\approx \mbox{arg}
\frac{(e^{i\alpha_3}\sqrt{(1-x_u)(1-x_d)}+e^{i\alpha_2}
\sqrt{x_ux_d})^*}{(e^{i\alpha_3}\sqrt{(1-x_u)x_d}-e^{i\alpha_2}\sqrt{x_u(1-x_d)})(e^{i\alpha_2}\sqrt{(1-x_u)x_d}-e^{i\alpha_3}\sqrt{x_u(1-x_d)})^*}\, ,\label{qpheexp}
\ee
for \(|(U_{CKM})_{13}| \ll 1 \), 
where we define
\bea
x_f&\equiv& \frac{d_f}{m_{f3}},\nonumber\\
\alpha_2&\equiv& \alpha_{u2}-\alpha_{d2}=\tau_u-\tau_d\equiv\Delta\tau,\\
\alpha_3&\equiv& \alpha_{u3}-\alpha_{d3}=\tau_u-\tau_d+(\sigma_u-\sigma_d)\equiv\Delta\tau+\Delta\sigma.\nonumber
\eea
We note that the values of \(m_b\) affects sensitively on the CKM quark mixings,
but the value of \(m_t\) does not.
\par
There are additional constraints among the parameters in the FZT model, when it is embedded in the SO(10) GUT model.
Let us first discuss the consraints in the quarks and charged leptons sector. 
In the SO(10) frame mentioned above, $S_f$ and $A_f$ ($f=u,d,e$) satisfy following relations:
\bea
4\alpha S_u &=& (3+\alpha \epsilon )S_d+(1-\alpha \epsilon )S_e,\\
\delta A_u &=& \delta^\prime A_d \, = \, \delta \delta^\prime A_e .
\eea
Expressing in terms of the components, we have the following constraints among the parameters:
\begin{align}
4\alpha a_u \cos(\Delta\tau+\tau_d) 
&= (3+\alpha \epsilon )a_d \cos\tau_d + (1-\alpha \epsilon )a_e \cos\tau_e,
\label{gut1}\\
4\alpha c_u \cos(\Delta\sigma+\sigma_d)
&= (3+\alpha \epsilon )c_d \cos\sigma_d+(1-\alpha \epsilon )c_e \cos\sigma_e,\\
4\alpha b_u 
&= (3+\alpha \epsilon )b_d +(1-\alpha \epsilon )b_e ,\\
4\alpha d_u &= (3+\alpha \epsilon )d_d +(1-\alpha \epsilon )d_e ,\\
\delta a_u \sin(\Delta\tau+\tau_d)
&= \delta^\prime a_d \sin \tau_d 
 = \delta \delta^\prime a_e \sin\tau_e,\\
\delta c_u \sin(\Delta\sigma+\sigma_d)&= \delta^\prime c_d \sin\sigma_d
 = \delta \delta^\prime c_e \sin\sigma_e.
\label{gut2}
\end{align}
From these constraints we have  
\begin{multline}
F(r,d_u,d_d)^2[4\alpha a_u\cos(\Delta \tau+\tau_d)-(2+\kappa)a_d\cos\tau_d]^2
-[4\alpha c_u\cos(\Delta\sigma+\sigma_d)-(3+\kappa)c_d\cos\sigma_d]^2  \\
=(1-\kappa)^2[a_e^2F(r,d_u,d_d)^2-c_e^2], \label{mat01}
\end{multline}
with constraints 
\begin{align}
-1&\le\cos\tau_e\equiv \frac{4\alpha a_u\cos(\Delta\tau+\tau_d)-(3+\kappa)a_d\cos\tau_d}{(1-\kappa)a_e}\le 1 ,\label{taue}\\
-1&\le\cos\sigma_e\equiv \frac{4\alpha c_u\cos(\Delta\sigma+\sigma_d)-(3+\kappa)c_d\cos\sigma_d}{(1-\kappa)c_e}\le 1 .
\label{sigmae}
\end{align}
where 
\begin{align}
\kappa &\equiv \alpha \epsilon, &
r & \equiv \frac{\delta'}{\delta}, &
F(r,d_u,d_d) & \equiv \frac{c_d\sin\sigma_d}{a_d\sin\tau_d}
 = \frac{c_e \sin \sigma_e}{a_e \sin \tau_e}=\frac{c_e \sin \sigma_e}{a_d \sin \tau_d}\delta.
\label{defF}
\end{align}
We note that  \(\tau_d\) and \(\sigma_d\) are obtained from Eqs.~(\ref{gut1})-(\ref{gut2}) as
\begin{align}
\tan\tau_d = \frac{a_u\sin\Delta \tau }
                   {r a_d -a_u \cos\Delta \tau } , \qquad
\tan\sigma_d = \frac{c_u\sin\Delta \sigma  }
                       {r c_d -c_u \cos\Delta \sigma  }. \label{tausigd}
\end{align}
And $\alpha$ and $\epsilon $ are determined as functions of $d_u$, $d_d$, and 
$d_e$:
\begin{align}
\alpha &= \frac{(m_d+m_s+m_b)d_e -(m_e+m_{\mu}+m_{\tau})d_d}
               {(m_d+m_s+m_b-m_e-m_{\mu}-m_{\tau})d_u-(m_u+m_c+m_t)(d_d-d_e)},\\
\kappa &= \frac{(m_u+m_c+m_t)(3d_d+d_e) -\left[3(m_d+m_s+m_b)+(m_e+m_{\mu}+m_{\tau})\right]d_u}
                         {(m_d+m_s+m_b-m_e-m_{\mu}-m_{\tau})d_u-(m_u+m_c+m_t)(d_d-d_e)}.
\end{align}
Note also that from Eq.~(\ref{defF}), when \(|\cos\tau_e|=1\), \(|\cos\sigma_e|\) becomes 1,
Therefore, in fact, we need only one of Eqs.(\ref{taue}) and (\ref{sigmae}).

As was shown in Eqs.~(\ref{CKMexp}) and (\ref{qpheexp}), 
the parameters $d_u$, $d_d$, $\Delta \tau $, and $\Delta \sigma $  are constrained 
by the observed CKM mixing matrix. 
The best fit is realized for the following values of the parameters,
\begin{eqnarray}
\Delta \tau  &=& \pi/2,\\
\Delta \sigma  &=& -0.121,\\
d_u &=& 0.9560\, m_t,\\
d_d &=& 0.9477\, m_b,
\end{eqnarray}
which are used in the following analysis.
Then $\tau_d$ and $\sigma_d$ are determined by two parameter $d_e$ and $r$ from Eqs.~(\ref{tausigd}) 
since the parameters $a_f,~b_f,~c_f$ are fixed by \(d_f\) from Eq.~(\ref{abc}). 
For these best-fit-parameters mentioned above, we obtain
\begin{eqnarray}
|(U_{CKM})_{12}|&=&0.2251,\\
|(U_{CKM})_{23}|&=&0.0340,\\
|(U_{CKM})_{13}|&=&0.0032,\\
\delta_q &=& 58.86^\circ . 
\end{eqnarray}
Here we have used the best fit values of the following quark and charged lepton masses 
estimated~\cite{Fusaoka} at the unification scale \(\mu=M_X\).
\begin{align}
m_u   (M_X)&=1.04^{+0.19}_{-0.20}\, \mbox{MeV},& 
m_d   (M_X)&=1.33^{+0.17}_{-0.19}\, \mbox{MeV}, \nonumber\\
m_c   (M_X)&=302^{+25}_{-27}\, \mbox{MeV}, &
m_s   (M_X)&=26.5^{+3.3}_{-3.7}\, \mbox{MeV}, \nonumber\\ 
m_t   (M_X)&=129^{+196}_{-40}\,  \mbox{GeV}, &
m_b   (M_X)&=1.00\pm0.04\, \mbox{GeV}, \label{eq123104} \\
m_e   (M_X)&=0.32502032\pm{0.00000009}\, \mbox{MeV}, &
m_\mu (M_X)&=68.59813\pm{0.00022}\,  \mbox{GeV}, \nonumber\\
m_\tau(M_X)&=1171.4\pm0.2\, \mbox{MeV}. \nonumber
\end{align}

Therefore we have two parameters $d_e$ and $r$ left to be fixed,
which  are determined from Eq.~(\ref{mat01}) with Eqs.~(\ref{taue}) and (\ref{sigmae}).
The solution of these equations is depicted in Figure 1 in the $d_e-r$ plane. 
As seen in Fig.~1 there are  two allowed curves which we call Sol.~(a) and Sol.~(b), respectively.
The typical values of $d_e/m_{\tau}$ for these solutions are 0.935883 for Sol.~(a) and 
0.307197 for Sol.~(b), which we shall use in the following analysis.

Thus we have succeeded to fit 13 parameters consistently in quark and charged lepton sectors. 
Using these parameters and $M_D$, $M_R$ and $M_L$, 
now let us proceed to discuss the neutrino masses and lepton mixings.
As seen from Eq.~(\ref{mass form}) we have three more free parameters $\delta''$,$\beta$, and $\gamma$ in the neutrino sector.
The neutrino mass matrix is given by the seesaw mechanism as
\begin{eqnarray}
M_\nu 
&= &M_L- M_D M_R^{-1} M_D^T .
\end{eqnarray}

\par
The diagonalization of the charged lepton mass matrix $M_e$ is simmilar to the quark mass matrix. 
On the other hand, since the mass matrix for the Majorana neutrinos is symmetric, 
$M_\nu$ is diagonalized as 
\begin{equation}
U_{\nu}^\dagger M_\nu U_{\nu}^* = \mbox{diag}\left(|m_{1}|, m_{2}, m_{3}\right),
\end{equation}
where $|m_{1}|, m_{2}, \mbox{and } m_{3}$ are real positive neutrino masses and  
the unitary matrix $U_{\nu}$ is described as
\begin{equation}
U_{\nu} =P_{\nu}^\dagger O_\nu Q_{\nu}.
\end{equation}
Here, in order to make the neutrino masses for the first generation to be real positive, 
we introduce an additional diagonal phase matrix $Q_{\nu}$ defined by 
\begin{equation}
Q_{\nu} \equiv \mbox{diag}\left(i,1,1\right). \ \label{Q}
\end{equation}
The MNS lepton mixing matrix $U_{MNS}$ 
of the model is given by 
\begin{equation}
U_{MNS}=U^\dagger_{Le}U_{\nu}=O^{T}_eP_{\ell}O_\nu Q_\nu,\label{MNS}
\end{equation}
where $P_{\ell} \equiv P_eP^\dagger_\nu$ is diagonal phase matrix. 
Eq.~(\ref{MNS}) is changed to the standard representation of the MNS 
lepton mixing matrix as well as the CKM quark mixing matrix, 
\begin{eqnarray}
U_{MNS}^{\rm std} &=& \mbox{diag}(e^{i\zeta_1^e},e^{i\zeta_2^e},e^{i\zeta_2^e})  
\ U_{MNS}  \nonumber \\
&=&
\left(
\begin{array}{ccc}
c_{13}^lc_{12}^l & c_{13}^ls_{12}^l & s_{13}^le^{-i\delta_l} \\
-c_{23}^ls_{12}^l-s_{23}^lc_{12}^ls_{13}^l e^{i\delta_l}
&c_{23}^lc_{12}^l-s_{23}^ls_{12}^ls_{13}^l e^{i\delta_l} 
&s_{23}^lc_{13}^l \\
s_{23}^ls_{12}^l-c_{23}^lc_{12}^ls_{13}^l e^{i\delta_l}
 & -s_{23}^lc_{12}^l-c_{23}^ls_{12}^ls_{13}^l e^{i\delta_l} 
& c_{23}^lc_{13}^l \\
\end{array}
\right) \nonumber \\
&&\times \mbox{diag}(1,e^{i \phi_2},e^{i \phi_3}) .
\label{stdrepLep}
\end{eqnarray}
Here \(\zeta_i^e\) comes from the rephasing in the charged-lepton fields,
\(\delta_\ell\) is the Dirac phase,
and \(\phi_i\) \(i=1,2\) are the Majorana phases in the MNS lepton mixing matrix.
\par
From Eq.~(\ref{MNS}) 
the neutrino oscillation angles and phases of the model are 
related to the elements as follows: 
\begin{eqnarray}
\tan^2\theta_{\mbox{{\tiny solar}}}& =&\frac{|(U_{MNS})_{12}|^2}{|(U_{MNS})_{11}|^2}\, ,\\
\sin^2 2\theta_{\mbox{{\tiny atm}}}& =&4|(U_{MNS})_{23}|^2|(U_{MNS})_{33}|^2\ . 
\end{eqnarray}
In Fig.~2 -- Fig.~5, we present the numerical values of the mixing angles in terms of the parameter $\beta \gamma $. 
Here we have used the following experimental constraints from global analysis of neutrino parameters \cite{strumia}.
\bea
0.25<\mbox{sin}^2\theta_{12}&<&0.38 \label{solmix},\\
0.35<\mbox{sin}^2\theta_{23}&<&0.65 \label{atmmix},\\
\mbox{sin}^2\theta_{13}&<&0.03\, . \label{chomix}
\eea
and
\bea
\Delta m_{21}^2&=&(7.2-8.9)\times 10^{-5} [eV^2] \label{solmass}, \\
\Delta m_{32}^2&=&(2.1-3.1)\times 10^{-3} [eV^2] \label{atmmass}\, .
\eea
Then the neutrino mass square ratio becomes
\begin{equation}
\Delta m_{21}^2/\Delta m_{32}^2 = 0.023 - 0.042.
\label{massratio}
\end{equation}
As is seen in Fig.~2 -- Fig.~5 (especially Fig.~4) we have consistent values of the lepton mixing angles with the observed experimental data if we take  $\delta^{\prime \prime}=0$ and tune $\beta 
\gamma$. Namely we obtain 
\begin{align}
\sin_{12}^2 &= 0.32, &
\sin_{23}^2 &= 0.52, &
\sin_{13}^2 &= 3.0\times 10^{-6}, &
\mbox{for } & \beta \gamma \simeq 3.4515\times 10^5  \mbox{ at Sol.~(a)},\\
\sin_{12}^2 &= 0.32, &
\sin_{23}^2 &= 0.53, &
\sin_{13}^2 &= 2.3\times 10^{-4}, &
\mbox{for } & \beta \gamma \simeq 2.9240 \times 10^7  \mbox{ at Sol.~(b)}.
\end{align}
On the other hand, for these parameters, the model predicts somehow small value for $\frac{\Delta m_{21}^2}{\Delta m_{32}^2}$, as seen in Fig.6.
\begin{align}
\frac{\Delta m_{21}^2}{\Delta m_{32}^2}&=1.7 \times 10^{-7} 
\mbox{\quad for Sol.~(a)}, &
\frac{\Delta m_{21}^2}{\Delta m_{32}^2}&=3.4 \times 10^{-4} 
\mbox{\quad for Sol.~(b).}
\end{align}

\section{Discussions}
In this paper we have considered the bridge between the phenomenological FZT model and SO(10) GUT. Namely we incorporated FZT in the framework of SO(10) GUT.
So our model suffers more stringent constraints than tha case when either only FZT or only SO(10) GUT was considered.
Neverthless, our model gave values consistent with the observations of quark-lepton. Only exception is the mass square ratio of neutrino.
However, this is not so bad news. For the other observables' fittings are very nice, and FZT
model is itself an approximation.
Also we have not considered the renormalization group equation (RGE) for neutrino sector. 
Namely we bottomed up the low energy spectrum to GUT and fixed the $M_R$
and could construct light neutrino mass matrix first at GUT.
So we must top down this matrix to the electroweak scale and fit with oscillation data \cite{Babu} \cite{f-o} at this scale. 
However, RGE effect works little on neutrino mixing angles and mass ratios if neutrino masses are not degenerate. 
Rather they depend on RGE of quark masses in which several ambiguities are left unsolved. Nobody knows clearly what is happening above TeV scale.
In most cases, the ratio \(m_{f1}/m_{f2}\) is stable 
but \(m_{f1}/m_{f3}\) is not.
Therefore, by changing only \(m_{f3}\) in the very wide range 
we will be able to check the stability of the bi-large mixing.

We note that we must check carefully when we change \(m_b\)
because the CKM quark mixing matrix is sensitive to \(m_b\).
Therefore only when we change the mass of bottom quark,
we use the following values:
\begin{itemize}
\item 
In the case \(m_b = (1.\mbox{GeV})\times 1/2\),
\begin{align}
\Delta\tau  &=  \pi/2, &
\Delta\sigma&= -0.314, &
d_u&=0.993m_t, &
d_d&=0.993m_b, 
\end{align}
\begin{align}
|(U_{CKM})_{12}| &= 0.2254,& 
|(U_{CKM})_{23}| &= 0.0347,& 
|(U_{CKM})_{13}| &= 0.0031,& 
\delta_q       &= 60.67^\circ.
\end{align}

\item
In the case \(m_b = (1.\mbox{GeV})\times 2\),
\begin{align}
\Delta\tau  &=  \pi/2, &
\Delta\sigma&= -0.063, &
d_u&=0.811m_t, &
d_d&=0.796m_b, 
\end{align}
\begin{align}
|(U_{CKM})_{12}| &= 0.2245,& 
|(U_{CKM})_{23}| &= 0.0327,& 
|(U_{CKM})_{13}| &= 0.0032,& 
\delta_q       &= 59.32^\circ.
\end{align}
\end{itemize}
We show, in Table 1 and Fig.7, how the lepton mixing angles and neutrino masses 
are sensitive to the quark and charged lepton masses of the third generation by changing factor 2 of them.
We find that RGE effects to our solution of the bi-large lepton mixing are small
from $m_t$ and $m_\tau$, while large from $m_b$.

\begin{table}
\begin{tabular}{c|cc|cc|cc} \hline
& \(m_t \times 2\) & \(m_t / 2\) 
& \(m_b \times 2\) & \(m_b / 2\) 
& \(m_\tau \times 2\) & \(m_\tau / 2\) \\ \hline
\(\beta\gamma\) 
& \(1.15 \times 10^8\)    & \(7.30 \times 10^6\)    
& \(8.61 \times 10^7\)    & \(7.60 \times 10^4\)
& \(1.24 \times 10^6\)    & \(2.95 \times 10^4\)\\
\(\sin^2\theta_{12}\)
& \(0.315\)               & \(0.333\)               
& \(0.311\)               & \(0.330\)
& \(0.312\)               & \(0.315\)\\
\(\sin^2\theta_{23}\) 
& \(0.573\)               & \(0.481\)               
& \(0.582\)               & \(0.460\)
& \(0.917\)               & \(0.131\)\\
\(\sin^2\theta_{13}\) 
& \(2.15 \times 10^{-4}\) & \(3.02 \times 10^{-4}\) 
& \(2.30 \times 10^{-4}\) & \(2.20 \times 10^{-4}\)
& \(1.55 \times 10^{-3}\) & \(2.25 \times 10^{-4}\)\\
\(\Delta m_{12}^2/\Delta m_{23}^2\)
& \(8.79 \times 10^{-4}\) & \(2.87 \times 10^{-4}\) 
& \(9.57 \times 10^{-4}\) & \(2.60 \times 10^{-4}\)
& \(1.07 \times 10^{-7}\) & \(1.77 \times 10^{-4}\) \\ \hline
\end{tabular}
\caption{The values around the peak when changing the third generaton masses.}
\end{table}



\begin{figure}[h]
\begin{center}
\end{center}

\caption{
The black lines satisfies Eq.~(\ref{mat01}).
The blue and red regions show the additional conditions 
Eqs.~(\ref{taue}) and (\ref{sigmae}), respectively.
The two overlapping regions of black line, red and blue areas in (i) are the allowed regions. 
The panels (ii) and (iii) are the enlarged figures of the allowed regions of (i) and are called Sol.~(a) and Sol.~(b), respectively, hereafter.} 
\end{figure}

\begin{figure}[h]
\begin{center}

\end{center}
\caption{
The Solution (a) in normal hierarchy. The red, blue, violet areas satisfy the 
solar Eq.~(60), atmospheric Eq.~(61), and CHOOZ Eq.~(62), respectively.
The green area shows the mass square ratio Eq.~(65).
(i) The diagram of $\beta\gamma$ versus $\delta''$.  (ii) The enlarged diagram of (i).
(iii) Mixing angles as functions of $\beta\gamma$: The solid red, blue, and violet 
lines depict sin$^2\theta_{12}$, sin$^2\theta_{23}$, and sin$^2\theta_{13}$, 
respectively.
}
\end{figure}

\begin{figure}[h]
\begin{center}
\end{center}

\caption{Same as Figure 2 but Sol.~(a) in inverted hierarchy. 
There is no red region in (i) and (ii) because sin$^2\theta_{12}\simeq 1$.
}

\end{figure}

\begin{figure}[h]
\begin{center}
\end{center}

\caption{Same as Figure 2 but Sol.~(b) in normal hierarchy. 
There is no violet region in (i) and (ii) because sin$^2\theta_{13}\ll 1$.
}

\end{figure}

\begin{figure}[h]
\begin{center}
\end{center}

\caption{Same as Figure 2 but Sol. (b) in inverted hierarchy.
There is no red region in (i) and (ii) because sin$^2 \theta_{12} \simeq 1$.}
\end{figure}

\begin{figure}[h]
\begin{center}
\end{center}
\caption{
The neutrino masses in Sol.~(a) and (b). 
The $m_i$ is the mass eigenvalues of light neutrinos. 
The red, blue, and green lines represent the neutrino masses in light order. }
\end{figure}

\begin{figure}[h]
\begin{center}
\end{center}

\caption{ The third generation mass dependence of the mixing angles as function of $\beta\gamma$.
The red, blue and violet lines show \(\sin^2 \theta_{12}\), 
\(\sin^2 \theta_{23}\) and \(\sin^2 \theta_{13}\), respectively 
in Sol.~(b) in normal hierarcy.
The top panel shows the values when we use the best values of masses 
in Eq.~(\ref{eq123104}).
Other panels show whether the bi-large mixing angle is reproduced or not
if the third generation mass would take double or half value.
Sol.~(a) has an behaviour similar to Sol.~(b)
}

\end{figure}

\end{document}